\documentclass[]{cupconf}
\usepackage{graphicx}

\title[Models of the Solar Vicinity]
      {Models of the Solar Vicinity: \\ The Metal Rich Stage}
      
\author[L.~ Carigi]{\\ Leticia Carigi }
\affiliation 
{Instituto de Astronom\'{\i}a, Universidad Nacional Aut\'onoma de M\'exico,
Apdo. Postal 70-264, M\'exico 04510 D.F., Mexico}

\begin{document}
\maketitle

\begin{abstract}
I present a review of chemical evolution models of the solar neighborhood.
I give special attention to the necessary ingredients to reproduce the observed [Xi/Fe] ratios 
in nearby metal and super metal rich stars, and to the chemical properties of the solar vicinity 
focusing on [Fe/H] $ \geq -0.1$. 
I suggest that the observed abundance trends are due to material synthesized and ejected by
intermediate mass stars with solar metallicity in the AGB stage,
and also by  massive stars with (super) solar metallicity in the stellar wind and supernovae stages.
The required tool to build chemical evolution models that reach super-solar metallicities is
the computation of stellar yields for stellar metallicities higher than the initial solar value.
Based on these models it might be possible to  estimate the importance
of merger events in the recent history of the Galactic disk
as well as the relevance of radial stellar migration from the inner to the outer regions of the Galaxy.
I also present a short review of the photospheric solar abundances
and their  relation with the initial solar abundances.
\end{abstract}

\firstsection

\section{Introduction}

The solar neighborhood is an invaluable laboratory for the chemical evolution models 
because the number of free parameters  is similar to the number of observational 
constraints.

A number of different assumptions are typically adopted by chemical evolution models
of a galactic zone: 
i) the galactic zone formation mechanism and formation time, 
ii)  when, how many, and what types of stars are formed,
iii) when those stars die, and, 
iv) which are  the chemical abundances of the material ejected during the life  
and death of the stars. 

Once a chemical evolution model for the solar vicinity satisfies the observational 
constraints, it is also possible to test both the Galaxy formation 
process and the properties of the underlying stellar populations. 
Therefore, the accuracy of the stellar and HII regions abundances estimations define 
the strength of our tests. 

For that reason, this review is in large extent  based on new data for the solar neighborhood:
i) HII regions obtained by Esteban, Peimbert and collaborators
(see Bresolin in this volume) and
ii)  F and G dwarf stars obtained by Bensby and Feltzing (see both of them  in this volume).
Also I will compare these data to the abundances of other galactic components and other galaxies, 
in particular to Bulge stars and extragalactic HII regions, 
in order to analyze the origin of the abundance trends at high metallicity.

\section {Observational Constraints}

The definition of ``solar vicinity" has several interpretations ranging from a 
zone including all low redshift galaxies until a region as small as the one including
the stars within 1 pc of the Sun. 
In the chemical evolution context the solar vicinity  corresponds  to a 
cylinder centered around the Sun, at 8 kpc from the Galactic center, that includes
objects belonging to the Galactic halo and disk. The dimensions of this cylinder depend on
the locations of the objects used as observational constraints. 
Typical adopted dimensions are 1 kpc in radius and $\sim$ 3 kpc in height, for a 
cylinder oriented orthogonally from the Galactic plane.

Since I am interesting in (super) metal rich objects,
I will  focus on  objects with metallicity near or higher than solar.

\subsection {Solar Abundances}

The photospheric solar abundances provide the reference pattern for general 
abundance determinations in the Universe (stars, ionized nebulae, galaxies) 
and, the inferred initial solar abundances correspond to the interstellar medium (ISM) 
of the solar neighborhood  4.5 Gyr ago. 
During the last $\sim$ 20 years the observational estimations of photospheric solar abundances 
regarding the most abundant heavy elements, like C, N, O, and Ne, have decreased their value. 

In Table 1 I show the chemical abundance determinations for some common elements 
in the solar photosphere computed by Anders \& Grevesse (1989, AG89), 
Grevesse \& Sauval (1998, GS98), and Asplund, Grevesse \& Sauval (2005, AGS05).
These abundances are in 12 + log(Xi/H) by number, 
I have added  the values of the mass fraction of He and metals, $Y$ and $Z$, respectively.
I also present the decreasing factors in the abundance determinations
between the Anders \& Grevesse work and the Asplund et al. data and 
between the Grevesse \& Sauval work and the Asplund et al. data. 

Since Fe is one of the most common elements and the value of its solar photospheric determination 
has kept almost constant during the last years, I will compare the available 
stellar data based on [Fe/H]. For HII regions, I will consider the O/H value
determined in those nebulae as the reference ratio, since  Fe is strongly dust depleted in 
ionized nebulae.

In order to reproduce the helioseismology observations, Sun models require more metals than 
the solar $Z$ value obtained by Asplund et al. (2005).
Bahcall et al. (2006) considering 
the photospheric metallicity of the Sun determined by Asplund et al. (2005)
found that the initial solar metallicity, $Z_{in}$, is 0.01405, 
15 \% more metals than those observed in the solar photosphere.
This difference is due to diffusive setting of the elements in the  photosphere 
during the last 4.5 Gyr, the age of the Sun
(for details, see Carigi \& Peimbert 2007).
This fact has implications for the chemical evolution
models, because the solar abundances in the photosphere have been taken as representative
of the abundances of ISM when the Sun was born, but those photospheric solar
abundances should be corrected by solar diffusion.
Moreover, the diffusive settling effect
should be considered in the abundance determinations of other stars,
taking into account that the amount of material settled depends on the stellar age. 

During this review I assumed that $Z_\odot=0.012$, $Z_{in}= 0.014$, and
$Z_{can}=0.020$ as the photospheric, initial, and canonical solar metallicity, respectively.

\begin{table}
\caption[]{Chemical Composition in the Solar Photosphere }
\begin{center}
\begin{tabular}{lcccrr}
\noalign{\smallskip}
Element & AG89  & GS98  & AGS05  &
$\frac {(Xi/H)_{AGS05}}{(Xi/H)_{AG89}}$ & $\frac {(Xi/H)_{AGS05}}{(Xi/H)_{GS98}}$  \\
 \hline
C  & 8.56 $\pm$ 0.04  & 8.52 $\pm$ 0.06 & 8.39 $\pm$ 0.05 & 0.68 & 0.74 \\
N  & 8.05 $\pm$ 0.04  & 7.92 $\pm$ 0.06 & 7.78 $\pm$ 0.06 & 0.54 & 0.72 \\
O  & 8.93 $\pm$ 0.04 & 8.83 $\pm$ 0.06 & 8.66 $\pm$ 0.05 & 0.54 & 0.68 \\
Ne & 8.09 $\pm$ 0.10  & 8.08 $\pm$ 0.06 & 7.84 $\pm$ 0.06 & 0.56 & 0.58 \\
Fe$^{a}$ & 7.51 $\pm$ 0.03  & 7.50 $\pm$ 0.01 & 7.45 $\pm$ 0.05 & 0.87 & 0.89 \\
$Y$ & 0.2743 & 0.2480 & 0.2486 & 0.91 & 1.00 \\
$Z$ & 0.0189 & 0.0170 & 0.0122 & 0.65 & 0.72 \\
\hline
\end{tabular}
\end{center}
$^{a}$Fe abundance in meteorites only for AG89.

\end{table}

\subsection {Abundances from HII Regions}

Abundance estimations in HII regions give us the preset-day abundances, for that 
reason they are very important chemical evolution models. 

Esteban et al. (2005) based on C and O recombination lines derived the C/H and O/H values of
8 Galactic HII regions. They found higher C/H and O/H values than the photospheric solar ones
for the Orion nebula and other 5 Galactic HII regions closer to the Galactic center (see Fig. 2).
These values are in agreement with the C/H  estimations derived  by 
Slavin \& Frish (2006, and references therein), who find C/H $= 8.78 \pm 0.20$,
a value higher than solar (photospheric and internal),
 along one line of sight of the Local Interstellar Cloud.  
HII regions gradients, particularly in the inner Galactic disk, might be useful to
analyze the chemical enrichment at high metallicities. 

Discussion on the abundances of  galactic and extragalactic metal rich HII regions, 
and also the different methods of abundances determinations
have been presented by Bresolin (2007 and this volume).
Since different methods to determine abundances provide different chemical abundances,
a consensus in the abundances determinations for HII regions is required.

\subsection {Age-[Fe/H] relation}

One of the most fundamental cosmochemistry observational results is the age-[Fe/H] relation 
because it links the age of the different dwarf stars with their chemical properties. 
This relation presents a large scatter along all of 
the metallicity range and metal rich stars are no exception. 
According to Bensby et al. (2005) thin disk stars with [Fe/H]$ > +0.2$ 
present ages between 3 and 9 Gyr,
but stars with $0 <$ [Fe/H]$ < +0.2$ have ages between 0 and 6 Gyr. 
According to Soubiran \& Girard (2006) the mean age of the thin disk stars of 
[Fe/H]$ > + 0.15$ is 5 Gyr with a dispersion of 3.4 Gyr, while the mean age of stars 
of  $0 < $[Fe/H]$ < +0.2$ is 3.8 with a dispersion of 2.1 Gyr. 
 
The dispersion is partly caused by stars born at other galactocentric radii with 
different star formation histories (SFHs) that  migrated  to the solar vicinity. 
Since this stellar migration requires time, Rocha-Pinto et al. (2006) showed that 
the solar neighborhood has been  polluted  by old and metal poor stars from inner and 
outer radii (between 6 and 9.5 kpc).
The age dispersion of the metal rich stars might be explained by a superposition of
young stars that were born in the solar vicinity and by old stars that were born at 
inner radii with an early and efficient star formation rate (SFR) as chemical evolution models of the 
Galactic disk predict (e.g. Carigi et al. 2005). 

An alternative explanation for the age dispersion presented in the age-[Fe/H] relation
are mergers of one or several satellite galaxies with different SFHs.

\subsection {[Xi/Fe] vs [Fe/H] relations} 
Other important observational constraints are provided by the [Xi/Fe] vs [Fe/H] relations 
derived by dwarf stars in the solar neighborhood. These relations 
give information to infer the past of the solar vicinity and the properties
of its stellar populations. 

{\bf $\bullet \ \alpha$ enhancement}

Some studies have found  $\alpha$ enhancement in thick disk stars 
compared to the thin disk stars in the  $ -0.7 < $ [Fe/H] $ < -0.1$  range  (Feltzing in this volume and
references therein).
No chemical evolution model that assumes a simple formation for the disk can 
reproduce that behavior. But, in the literature there are some models with complex disk formation
histories that can explain the $\alpha$ enhancement:

i) Nykytyuk \& Mishenina (2006) suggest a two-zone model with different gas infalls and SFHs
 for the thick and the thin disks. 
Their model can reproduce the $\alpha$ enhancement in the thick disk stars compared to the thin disk stars, 
but not the  dispersion in the age - [Fe/H] relation.
Chiappini (2001 and this meeting) has suggested a similar model assuming a double infall for
the Galactic disk and the thin disk formed with material from the thick disk and from the intergalactic medium.

ii) Brook et al. (2005) suggest hierarchical mergers and fragmentation models. 
Specifically the thick disk formed by multiple gas rich mergers at early times (7.7 Gyr ago) 
at redshifts higher than $\sim$ 1.
Part of the gas of the thick disk was leftover by shock heating, 
then the thin disk formed from primordial infalling gas 
and the pre-enriched gas by the thick disk stars that falls later onto the thin disk.
Their results are in good agreement with $\alpha$ enhancement but in partial agreement
with the dispersion in age-[Fe/H] relation.

{\bf $\bullet \ $ Metal Rich Disk Stars}

Bensby et al. (2005) have determined chemical abundances in F and G dwarfs
and they found strong abundance trends, which are shown in Fig. 1. 
As can be noted from this figure
 the slope of [Xi/Fe] vs [Fe/H] relation for [Fe/H] $> - 0.1$:
i) change significantly for  Na, Ni, and Zn, with $\Delta [Xi/H]/\Delta [Fe/H] > 0 $,
and for Ba with $\Delta [Xi/H]/\Delta [Fe/H] < 0 $,
 i) change moderately for Mg, Al, Si, Ca, and Ti, being $\Delta [Xi/H]/\Delta [Fe/H] \sim 0 $, and
iii) does not change for O, Cr, and Eu.

Similar trends have been observed in the red giants of the Galactic Bulge 
(Cunha \& Smith 2006, Leceurer et al. 2006):
i) the [Na/Fe] increase with [Fe/H],
ii) the [Na/Mg] increase with [Fe/H] $> 0$ but not as much as in the thin disk, and
iii) the [O/Mg] decrease with [Fe/H] $ > 0$ like in the disk stars.

Recently Johnson et al. (2006) have determined  abundances for a super metal rich 
G-dwarf of the Galactic bulge and they find that the [alpha/Fe] ratios are subsolar, 
while the odd-Z elements are slightly supersolar, 
these values are in agreement with the trends seen in the more metal-rich stars of the Galactic disk
(see Fig. 1).

\begin{figure}
\includegraphics[width=5.5in,height=6.4in]{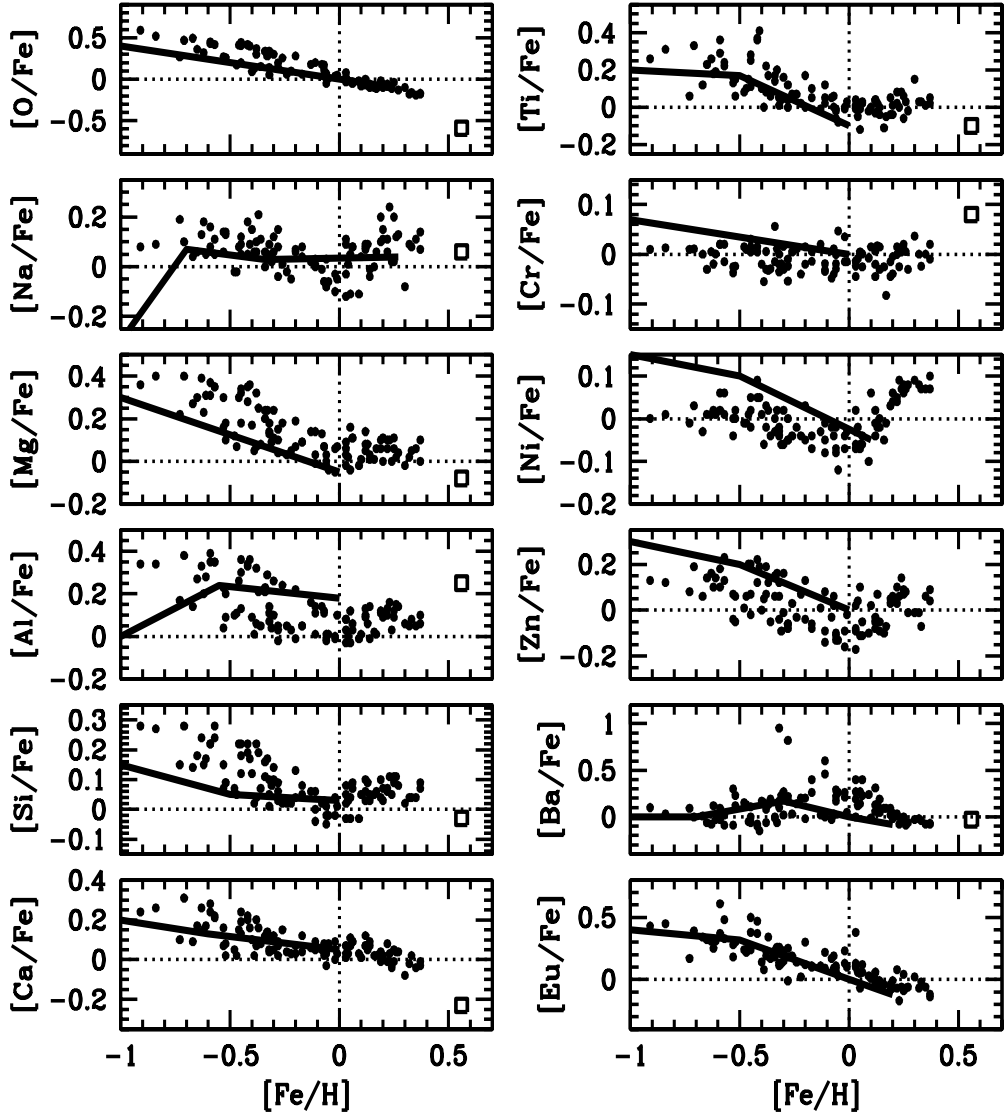}
\hfill
\caption{Evolution of [Xi/Fe] vs [Fe/H] predicted by different models:
{\it Al} by Timmes et al. (1995),
{\it Ca} by Portinari et al. (1998),
{\it Ti, Ni} and {\it Zn} by Francois et al. (2004),
{\it O} by Gavil\'an et al. (2005),
{\it Mg, Si} and {\it Cr} by Prantzos (2005),
{\it Ba} and {\it Eu} by Cescutti et al. (2006) and
{\it Na} by Izzard et al. (2006).
{\it Filled circles:} F and G dwarf disk stars by Bensby et al. (2005).
{\it Open squares:} The most metal rich G dwarf Bulge star by Johnson et al. (2006).
[Xi/H] corrections due to different photometric solar values assumed by the authors are not made. 
}\label{fig1}
\end{figure}

\section{Chemical Evolution Models of the solar vicinity}

The goal of any chemical evolution model is to explain the observed chemical properties.
In the literature there are successful models 
that match the abundance trends for [Fe/H] $ \leq  0$,
for example, see the excellent reviews by Gibson et al. (2003) and by  Matteucci (2004).
Those models are computed with different codes and  assumptions.
In Fig. 1 I present the [Xi/Fe] vs [Fe/H] evolution predicted by some well known models.
If those theoretical trends are extrapolated to [Fe/H] $\sim + 0.5$
no chemical evolution model is able reproduce the change in the [Xi/Fe] vs [Fe/H] relation
for [Fe/H] $ > -0.1$.

Therefore, I will study the dependence of [Xi/Fe] with the different
ingredients of a chemical evolution model,
in order to find an explanation of the  [Xi/Fe] trends presented by metal rich stars
of the Galactic thin disk.

It is well known that the [Xi/Fe] ratios depend on gas flows, star formation rates,
 initial mass function, and stellar yields,
being the last two ones the most important factors.

\subsection {Gas and star flows}

Infalls and outflows change the [Xi/Fe] ratios depending on the abundances and
the amount of gas of the flows.
Rich outflows with SNII material reduce [Xi/Fe] as opposed to the increment  required
by most of the observed trends.
Rich outflows  with SNIa material increase [Xi/Fe] but they also  decrease [Fe/H]
preventing  the formation of  metal rich stars.
A metal rich gas infall of  overabundant elements present in metal rich stars, like Na,
can reproduce the [Xi/Fe] raise, but how does the infalling gas get those [Xi/Fe] values?

Brook et al. (2005) explain the chemical properties of the  thick and thin disk stars with
models that assume mergers and infalls mainly at redshifts lower than 1,
but their results do not predict the abundance trends for $[Fe/H] > -0.1$.

Reddy (in this meeting) showed a secondary peak in that [Fe/H] distribution for [Fe/H] $ > 0$.
An inclusion of a significant amount of stars (or gas that triggered the star formation)
from a merger event could explain  the secondary peak.
If the thin disk metal rich stars formed in one o several galactic satellites
that settled in the Galactic disk,
how did the stars of those satellites reach supersolar [Fe/H] with (super)solar [Xi/Fe] ?

Based on the merger scenario, the Bulge also formed by satellites that fell early in the Milky Way,
therefore the origin of old and metal rich stars of the Bulge and Galactic disk is in small galaxies
or metal rich stars form of the material comes from small structures, but again,
how did those structures reach [Xi/Fe]$ > 0$ ?

Based on another scenario the Bulge could form by the stars that were born in the inner 
Galactic disk and were dynamically heating by the bar (Col\'{\i}n et al. 2006).
Moreover, the same bar could be able to produce radial flows of stars
from the inner to the outer part of the Galactic disk. 
Therefore the metal rich stars of the Bulge and the solar neighborhood formed in the inner disk,
but how did the inner disk reach [Xi/Fe]$ > 0$ ?

Radial gradients can be powerful tools 
to decide if metal rich stars observed in the solar vicinity and the Bulge 
formed in situ or alternatively were formed in inner galactocentric radius or 
merged satellites. 
In the most complicated case (or the most realistic) a combination of these three ones 
should be the answer.
 
Therefore, stellar or gaseous infalls can explain the abundance trends observed 
for $[Fe/H] > -0.1$ but they pose the question of how could these infalls get those 
(super) solar [Xi/Fe] values?

\subsection  {Star formation rate}

Important changes in the star formation rate,  affect the  [Xi/Fe] ratios
mainly after a significant star formation burst
(e.g. Carigi et al. 1999, 2002; Chiappini  2001).

The spiral wave is the most important inner mechanism that triggers  star formation
and that  recently could have formed stars from a metal-rich gas.
Rocha-Pinto et al. (2000) and Hern\'andez et al. (2000) inferred the star formation history 
 from the color magnitude diagram for the solar vicinity.
They found: 
i) a decreasing exponential general behavior of the SFH in the last $\sim$ 10 Gyr, and
ii) variations from the general behavior related directly to the spiral wave passages.
Based on these results, it is found that there were no significant bursts of star formation 
in the last $\sim$ 6 Gyrs,
therefore it is unlikely that a burst could have modified the [Xi/Fe] slope.

An important fact is that  metal rich stars with similar [Xi/Fe] values
have been observed in the solar neighborhood and in the Bulge,
galactic components with different SFHs.
The solar vicinity had a moderate SFR during 12 Gyr,
(e.g. Carigi et al. 2005)
while the Bulge formed very quickly, in less than 0.5 Gyr, with a high SFR
(Ballero et al. and Matteucci, both in this volume).

Therefore, I discard changes in the star formation rate as the explanation of the change in the [Xi/Fe] slopes
for  [Fe/H] $> -0.1$.

\subsection {Initial Mass Function}

The initial mass function, IMF, gives the mass distribution of the formed stars in a 
star formation burst.
This function is parametrized by the slope for different mass ranges and by
the lower and upper mass limits of the formed stars.
Since [Xi/Fe] depends strongly on the IMF, 
a dependence of the IMF with metallicity, density or gas mass
could explain the change of the [Xi/Fe] slope.

According to Kroupa (in this volume) there is no evidence that the IMF
changes with $Z$.
Nevertheless, Bonnell suggested (in this meeting) that the IMF changes with $Z$:
for supersolar metallicities the slope for massive stars (MS)  could be steeper and 
the upper limit could be lower than for  subsolar metallicities, producing 
[Xi/Fe] subsolar values, in contradiction with the observed values.

In a metal rich gas it is more difficult to create MS due to
the Jeans  mass dependence on $Z^{-2/3}$ and metal rich stars truncate the star formation
process by their stellar winds.
This suggestion could explain the low ionization of the metal rich HII regions
compared to that of  HII regions with subsolar metallicity:
in a metal rich gas the number of MS may be lower leading to a smaller
number of ionizing photons,
on the other hand, the lower stellar temperatures of the metal rich stars help to 
to reduce the number of ionizing photons, further work on this suggestion needs to be done.

The IMF dependence on the gas density should be less important,
because the same abundance trends have been observed in Galactic components
with different densities (Bulge, open clusters, isolated dwarfs),
but with a same property: super solar metallicity.

According to Weidner \& Kroupa (2005) the IMF  depends on gas mass.
They found that the slope and the upper mass limit change with gas mass available to form stars.
In dwarf galaxies the upper mass limit is lower 
and the slope in the MS range is steeper
producing lower [Xi/Fe] values than those of  normal galaxies
for elements synthesized only by MS. 

Moreover, Carigi \& Hern\'andez (2007) found important effects on the abundance ratios 
when the IMF is  stochastically populated.
The [O/Fe] values varied within three orders of magnitude
for a stellar population of  500 M$_\odot$ that enrich  a gas mass
of  $10^4$ M$_\odot$.
This effect could  explain the dispersion observed in the abundances ratios,
but not the abundance trends.

Therefore, possible modifications in the initial mass function cannot explain the abundance trends observed 
for  [Fe/H] $> -0.1$.

\subsection {Stellar Yields}

\begin{figure}
\includegraphics[width=5.0in,height=5.0in]{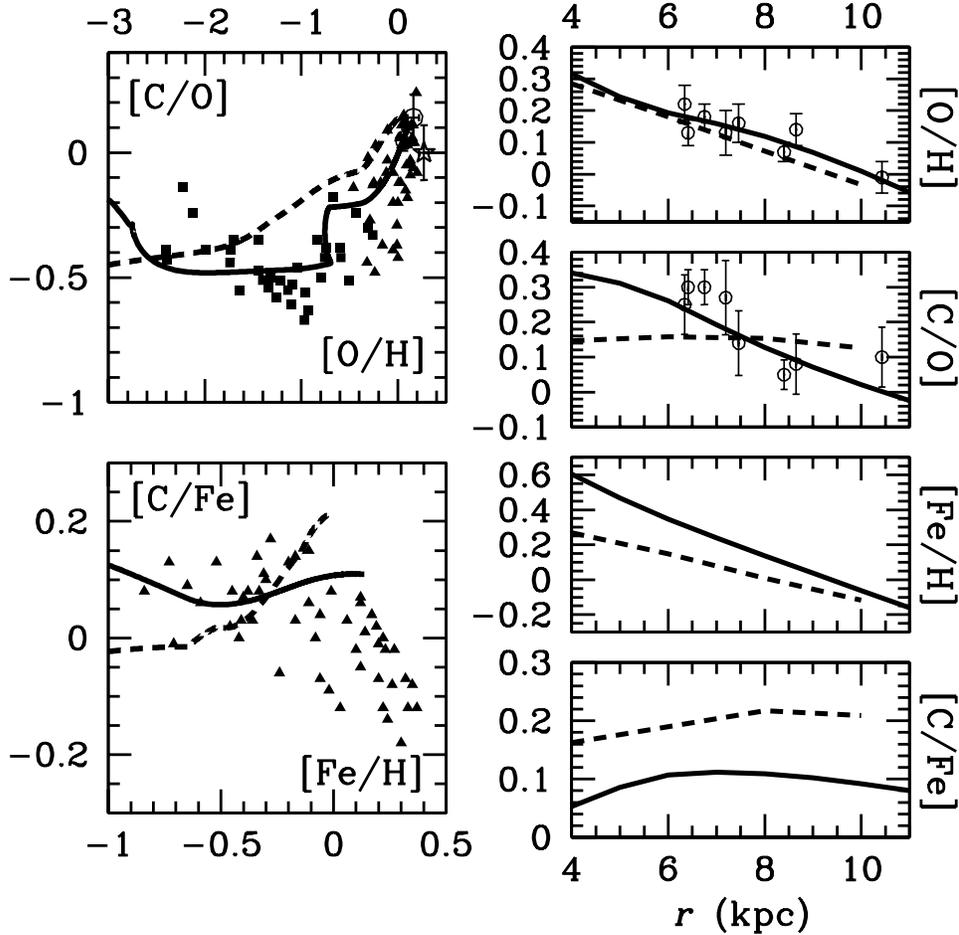}
\hfill
\caption{
Model predictions by Carigi (2000) ({\it dashed lines})
 and Carigi et al. (2005)  ({\it continuous lines}) considering
yields of metal-rich massive stars by Portinari et al. (1998) and Maeder (1992), respectively.
{\it Left Panels:} [C/O,Fe] evolution with [O,Fe/H]  in the solar vicinity.
{\it Right Panels:} Present-day ISM abundances ratios as a function of galactocentric distance.
{\it Open circles:} Galactic HII regions, gas plus dust, by Esteban et al. (2005) and 
Carigi et al. (2005).
{\it Star:} Extragalactic HII region (H1013) in M101 by Bresolin (2007).
{\it Filled triangles:} F and G dwarf disk stars by Bensby \& Feltzing (2006).
{\it Filled squares:} dwarf stars by Akerman et al. (2004).
Photometric solar values by AGS05 are considered except for data by Bensby \& Feltzing (2006)
because they assumed their own solar abundances. 
}\label{fig2}
\end{figure}

Since supersolar [Xi/Fe] values seem to be a common property of stars with [Fe/H]$ > 0$
in  galactic components (thin disk, bulge, open clusters)
with different formation histories,
the abundance trends can be explained due to 
the stellar yields of (super)solar metallicity stars.
The observed abundances will provide strong constraints on the physical processes taking place
in the stellar cores.

The models shown in the Fig. 1 consider different $Z$-dependent yields for massive stars,
for low-and-intermediate mass stars (LIMS), and for SNIa.
These stellar yields were computed for $Z \leq Z_{can}$ with the exception
of the Portinari et al. (1998) yields, but these authors never used their yields
for $Z = 0.05$ because they stopped their computations at [Fe/H] $ = 0 $.
Cescutti et al. (2006) and Fran\c{c}ois et al. (2004) modified the stellar yields
obtained by stellar evolution models in order to reproduce the observed trends
for [Fe/H] $< +0.1$.

Edmunds (in this meeting) suggested that stellar yields increasing with $Z$
raise the  [Xi/Fe] values for  [Fe/H]$ > -0.1$.
The [O, Mg/Fe] values for Bulge and thin disk stars
indicate a $Z$ dependence in the ratio of the O to Mg yield.

Meynet et al. (in this volume) show that no-rotating MS with  $Z = Z_{can}$
and a high mass loss rate eject more C than O,
and that these stars are an important source of  He, C, Ne, and Al, but not so much of O.
These facts could explain the [O/Fe]  decrease with increasing [Fe/H]
while the [Al/Fe] values remain almost constant for  [Fe/H] $ \ge 0$.

The significant change in the [Na/Fe] slope suggests an extra source of Na production.
Assuming that SNII and AGB stars
produce Na, Izzard et al. (2006) reproduce the [Na/Fe] values for [Fe/H] $< - 0.2$,
but fail to reproduce the [Na/Fe] increase
for $[Fe/H] > 0$. They suggest that the change in the [Na/Fe] slope may be explained
by secondary Na produced by SNII.
Nevertheless, according to  Frohlich (in this meeting) the core collapse supernovae
cannot explain the [Na/Fe] increase observed for  [Fe/H] $ \ge 0$.

Another channel that contributes to the enrichment of a metal rich gas is provided by
SNIa.
According to Yoon (in this volume) there are different scenarios for SNIa
with different time delays, but the amount of heavy elements ejected is similar for
the different scenarios.
The role of rotation might be important in the production of chemical
elements, but this effect has not been  studied yet.

Therefore, new stellar yields for massive stars and intermediate mass stars of
solar and supersolar metallicity  are required.

{\bf $\bullet \ $ Importance of Stellar Winds in Metal Rich Stars}

One of the most important problems in the chemical evolution of  Galaxies is the C production.
Carigi et al. (2005, 2006)
have studied the contribution of the C enrichment
by MS and LIMS in different types of galaxies.
We have found that the MS have contributed with 48 \% and 36 \%
to the total C produced in the solar neighborhood and in the dIrr galaxy NGC 6822, respectively.
The difference is due to the $Z$  effect on the stellar winds of MS.
Massive stars of solar $Z$ eject more C than those of subsolar $Z$
through stellar winds (see Meynet et al. and Crowther, both in this volume). 

Carigi et al. (2005) made a chemical evolution model of the Galaxy where they
assumed that the metal rich stars behave like stars of $Z_{can}$.
They are able to reproduce the C/O 
and O/H values  of the solar vicinity as well as the O/H and C/O gradients observed
by Esteban et al. (2005) but cannot reproduce the decrease in the [C/Fe] for [Fe/H] $> -0.1$ 
shown by Bensby \& Feltzing (2006) and Allende-Prieto in this meeting (see Fig. 2).

Carigi (2000) made a model considering yields of MS by Portinari et al. (1998)
for $Z = 2.5 Z_{can}$ and predicted a C/O gradient flatter than the observed one.
The flattening of the gradient is due to the increase in 
the mass-loss rate with $Z$  ($\propto Z^{0.5}$) assumed by Portinari et al., 
consequently metal-rich MS are stripped
before C is synthesized and their C yields are lower than those of stars with $Z = Z_{can}$.

Based on Meynet et al. (in this volume)
the rotating stars of $Z = Z_{can}$ are more efficient ejecting C and O
than the rotating stars of  $Z = 2 Z_{can}$.
This could explain the [C/Fe] decrease shown by metal rich stars of the thin disk,
but  the C contribution due to  LIMS must be included also to have a complete picture of
the evolution at high $Z$.

In order to reproduce the high C/O values for inner galactocentric radii 
the mass-loss rate for metal-rich stars has to be lower than that assumed by Portinari et al. (1998).
On the other hand, to reproduce the low C/Fe values for [Fe/H] $> 0 $ in the solar neighborhood
the mass-loss rate has to be higher than that assumed by Portinari et al. 
Puls (in this meeting) gave limits for the mass-loss rate, that depends as $Z^{0.62\pm0.15}$.

Consequently, there is an inconsistency between theory and observations for the
behavior of  C/O and C/Fe for high metallicities
and a more complex explanation is needed.

\section{Conclusions}

Chemical evolution models of the solar vicinity for metal rich stars are in the early stages
of development, nevertheless based on this review I present the following  conclusions:

\begin{itemize}

\item  
Models that assume hierarchical mergers and fragmentation explain most of the chemical and
kinematic properties of thick and thin disks for [Fe/H] $ \leq  0$.

\item 
Models that assume different star formation histories and infalls for the thin and the thick disk 
explain only their chemical properties  for [Fe/H]$ \leq  0$.

\item
There is no published chemical evolution model of the solar vicinity that can reach
the maximum [Fe/H] value observed in the thin disk, that is, [Fe/H] $\sim  + 0.4$.

\item 
Simple extrapolations of chemical evolution models that assume [Fe/H] $ \le 0 $ to the 
[Fe/H] $\sim  + 0.4$  regime,  result in predictions that fail to match the chemical abundances 
found in (super) metal-rich stars. 

\item
The similar abundance ratios trends observed for stars in the solar neighborhood and 
the high metallicity Bulge stars suggest that both were created by  the pollution of 
supersolar massive stars and solar intermediate mass stars. 
The stellar yields of these stars are required to study the metal enrichment of the
interstellar medium in the solar vicinity and the Bulge.

\item
No current chemical evolution model includes adequately the contribution of
 metal-rich massive stars, because their yields have not been computed completely yet.

\item
The  intermediate mass stars in the AGB may produce an important amount of the 
chemical elements heavier than oxygen, but in the case of Na their contribution 
is not enough to explain the [Na/Fe] rise for [Fe/H]$ > - 0.1$. 

\item 
The large dispersion in age shown by metal rich stars in the solar vicinity could be indicating
an external origin: like merger events or stellar migrations from the inner disk. 

\end{itemize}

\smallskip 
{\it In my opinion, we are going into a new phase of Astronomy: 
from the past to the future of the Universe
(before this meeting the emphasis was from the present to the past)
the metal-rich past and present give us hints to the future}.

\medskip
I thank Garik Israelian and the Organizing Committee for kindly inviting me to give this review.
I am grateful to Manuel Peimbert for several fruitful discussions and helpful suggestions,
to Brad Gibson for sending me the paper by Izzard et al. in advance of publication, and to
Octavio Valenzuela for a careful reading of the manuscript.
I received partial support from the Spanish MCyT under project AYA2004-07466 to attend this meeting.

\end{document}